\documentclass[pra,twocolumn,showpacs,preprintnumbers,amsmath,amssymb,superscriptaddress]{revtex4-1}

\usepackage{graphicx}
\usepackage{dcolumn}
\usepackage{bm}
\usepackage{color}
\usepackage{amsmath}
\usepackage{subfig}
\usepackage{verbatim}

\newcommand\ot{\otimes}

\renewcommand{\>}{\rangle}

\def\duzomniejsze{<\kern-.7mm<}
\def\duzowieksze{>\kern-.7mm>}
\def\blacksquare{\vrule height 4pt width 3pt depth2pt}

\def\textbf#1{{\bf #1}}
\def\beq{\begin{equation}}
\def\eeq{\end{equation}}
\def\be{\begin{equation}}
\def\ee{\end{equation}}
\def\ben{\begin{eqnarray}}
\def\een{\end{eqnarray}}
\def\beqa{\begin{eqnarray}}
\def\eeqa{\end{eqnarray}}
\def\eea{\end{array}}
\def\bea{\begin{array}}
\newcommand{\bei}{\begin{itemize}}
\newcommand{\eei}{\end{itemize}}
\newcommand{\bee}{\begin{enumerate}}
\newcommand{\eee}{\end{enumerate}}

\newtheorem{lemma}{Lemma}
\newtheorem{corollary}{Corollary}
\newtheorem{proposition}{Proposition}
\newtheorem{theorem}{Theorem}
\newtheorem{definition}{Definition}
\newtheorem{observation}{Observation}

\newtheorem{rem}{Remark}

\newcommand{\bia}[6]{#1, #3 {\bf #4}, #5 (#6).}

\def\A{{\mathcal{A}}}
\def\B{{\mathcal{B}}}
\def\E{{\mathcal{E}}}
\def\F{{\mathcal{F}}}
\def\I{{\mathcal{I}}}
\def\M{{\mathcal{M}}}
\def\R{{\mathcal{R}}}
\def\S{{\mathcal{S}}}
\def\T{{\mathcal{T}}}
\def\LL{{\mathcal{L}}}
\def\mix{{\textrm{mix}}}
\def\uu{{\textrm{u}}}

\begin{document}


\title{Axiomatic approach to contextuality and nonlocality}

\author{Karol Horodecki}
\affiliation{National Quantum Information Center of Gda\'nsk, 81-824 Sopot, Poland}
\affiliation{Institute of Informatics, University of Gda\'nsk, 80-952 Gda\'nsk, Poland}

\author{Andrzej Grudka}
\affiliation{Faculty of Physics, Adam Mickiewicz University, 61-614 Pozna\'n, Poland}

\author{Pankaj Joshi}
\affiliation{National Quantum Information Center of Gda\'nsk, 81-824 Sopot, Poland}
\affiliation{Institute of Theoretical Physics and Astrophysics, University of Gda\'nsk, Gda\'nsk, Poland}

\author{Waldemar K\l{}obus}
\affiliation{Faculty of Physics, Adam Mickiewicz University, 61-614 Pozna\'n, Poland}

\author{Justyna \L{}odyga}
\affiliation{Faculty of Physics, Adam Mickiewicz University, 61-614 Pozna\'n, Poland}




\date{\today}

\begin{abstract}

We present a unified axiomatic approach to contextuality and non-locality based on the fact that both are resource theories. In those theories the main objects are consistent boxes, which can be transformed by certain operations to achieve certain tasks. The amount of resource is quantified by appropriate measures of the resource. Following recent paper [J.I. de Vicente, J. Phys. A: Math. Theor. {\bf 47}, 424017 (2014)], and recent development of abstract approach to resource theories, such as entanglement theory, we propose axioms and welcome properties for operations and measures of resources. As one of the axioms of the measure we propose the asymptotic continuity: the measure should not differ on boxes that are close to each other by more than the distance with a factor depending logarithmically on the dimension of the boxes. We prove that relative entropy of contextuality is asymptotically continuous. Considering another concept from entanglement theory---the convex roof of a measure---we prove that for some non-local and contextual polytopes, the relative entropy of a resource is upper bounded up to a constant factor by the cost of the resource. Finally, we prove that providing a measure $X$ of resource does not increase under allowed class of operations, such as e.g. wirings, the maximal distillable resource which can be obtained by these operations is bounded from above by the value of $X$ up to a constant factor. We show explicitly which axioms are used in the proofs of presented results, so that analogous results may remain true in other resource theories with analogous axioms.
We also make use of the known distillation protocol of bipartite nonlocality to show how contextual resources can be distilled.

\end{abstract}


\maketitle

\section{Introduction}

Quantum contextuality stands among the most expressive manifestations of nonclassicality in quantum mechanics \cite{KS67,Bell66}.
In recent years it has attracted much attention and has been a topic of extensive studies \cite{con1,con2,con3,con4}. Apart from the interest focused on fundamental concepts, quantum contextuality has been associated with fast computing \cite{raus} and quantum information processing \cite{CAEGCXL14}, which opens a path to possible application of contextual resources in different scenarios.

A particular example of contextuality in the framework where two or more spatially separated parties perform measurements on each subsystems has been termed nonlocality \cite{BellNL}. A lot of effort has been devoted to classifying and quantifying the nonlocality which was identified as a useful resource in the device independent quantum information processing (see also Sec. IV of Ref. \cite{BellNL}. Although different in nature, nonlocal correlations as well as quantum entanglement proved useful in information processing tasks which cannot be performed with the sole use of classical correlations. This in turn led to the development of the resource theories of entanglement \cite{Hor4} and nonlocality \cite{Vin14}.

While approaching to the formulation of a resource theory three basic ingredients need to be considered. Firstly, the concept of a resource needs to be developed, and showed that it is {\it useful} regarding some specific tasks, which remain unattainable while having only non-resource objects at disposal. Secondly, there must be operations by which one may transform resources into one another. Thirdly, one needs to have a tools to compare different objects by means of measuring the resource contained by them, namely a measure of the resource.

Only recently a theory of resources has been formalized with respect to nonlocal resources \cite{Vin14}, steering resources \cite{ressteer}, as well as a general abstract characterization of resource theories has been formulated \cite{fryc}, which captures all needed features and relations that they have in common. In this light we develop the recent theory of nonlocal resources \cite{Vin14} to include the notion of contextuality the particular manifestation of which is nonlocality. After identifying the contextual system as useful regarding some computational tasks \cite{nat}, in this paper we treat contextual systems (``boxes'') as resources in similar way as it has been done with respect to nonlocal resources. In particular, we describe the notion of contextuality and then, based on the resource theory of entanglement, we formulate a set of axioms for the transformations of contextual resources, as well as for measures intended to quantify the value of given resources.

One of the axioms that we explore is asymptotic continuity --- a very desired property from the experimental point of view. Suppose we wish to quantify the amount of resource of a box $B$ using the measure $X$. However, the experimental realization of the box produces an imperfect box $B'$ which is possibly close to $B$:
\ben
||B - B'|| \leq \epsilon,
\een
where $||.||$ denotes the {\it trace distance between two boxes} \cite{CT}. For any measure of a resource we want it to differ not more than the distance between two considered boxes times a constant logarithmic factor of the dimensionality of the box.
Asymptotic continuity of a measure $X$ therefore means that
\ben
|X(B) - X(B') | \leq \epsilon \log d + f(\epsilon),
\een
where $d$ is dimensionality of the boxes and $f$ is a function such that $f(\epsilon) \xrightarrow{\epsilon \rightarrow 0} 0$.

Only recently the measures of contextuality has been developed, which enables to quantify the amount of contextuality of boxes \cite{context-measures}. In particular two measures of contextuality has been introduced: \textit{mutual information of contextuality} (MIC) and \textit{relative entropy of contextuality} (REC). As stated earlier, from experimental point of view it is mandatory that any measure of a resource, in particular the measure of contextuality is asymptotically continuous.
With this respect the main result of the paper is proving that the measure MIC fulfills the axiom of asymptotic continuity. Since the measures MIC and REC are equivalent \cite{context-measures}, the asymptotic continuity holds also for the latter, relative entropy distance measure.

The next result of the paper is showing that the relative entropy measure of a resource is upper-bounded (up to a constant factor) by another measure which has been defined in Ref. \cite{context-measures}, the cost of a resource. We then give examples of applications of this result for a class of bipartite boxes with binary inputs and binary outputs (the most nonlocal of which is $PR$-box \cite{Rastall85,PR94}), as well as a class of boxes related to contextual $n$-cycles \cite{AQBCC13}.
Furthermore, we consider distillation of a resource with the regard that the measure of a resource fulfills the axiom of monotonicity (i.e. that the measure does not increase under the set of allowed operations), and show that distillable contextuality is upper-bounded (up to a constant factor) by the value of the measure.
We also make use of a distillation protocol as originally devised in \cite{dist-nonl} to show how contextual resources can be distilled.
We then analyze the application of the bound with respect to two relative entropy-based measures of contextuality.

\section{Axiomatic approach to resource theory of contextuality and nonlocality}

We present a framework for a construction of a general resource theory of contextuality, which can be regarded as a development of resource theory of nonlocality as presented in Ref. \cite{Vin14}. After formalizing the notion of contextual resources with the relation to nonlocal  resources, which are to be understood as a specific form of the former, we proceed with formulating the axioms for operations on contextual resources, and axioms for the measures intended to quantify the value of a given resource.

\subsection{Contextual or nonlocal boxes}

In the present setting the object of interest are the measurement statistics, without any references to what actual measurements are being performed on actual physical systems. At this point we do not need to assume that the measurements are performed by spatially separated parties, which in fact the notion of nonlocality is all about.
In the present view we consider a set of observables $\mathcal{M}$ that can be performed on any physical system, where a measurement of each $M_i \in \mathcal{M}$ gives an outcome $m_i$ with a probability $p(m_i|M_i)$. Let us assume that there exist subsets $\mathcal{M}_i$ of jointly measurable observables. Each such subset we will call a \textit{context} denoted by $c$. The joint probability distribution of obtaining the outcomes $(m_{1},m_2,...,m_k)$ while measuring the observables $M_1,M_2,...,M_k$ belonging to a context $c$ we will denote by
\ben
p(\lambda_c):=p(m_{1},m_2,...,m_k|M_{1},M_2,...,M_k),
\een
where $\lambda_c = (m_{1},m_2,...,m_k)$ such that $M_{1},M_2,...,M_k \in c$.
Note that an observable $M_i$ may belong to several different contexts.
A \textit{box} ($B$) is then a set of joint probability distributions $B=\{p_B(\lambda_c)\}$ for all contexts in $\mathcal{M}$ (we will omit the subscript $B$ of $p_B(\lambda_c)$ when it is not necessary).

There is, however, a significant constraint which must be obeyed by all boxes to be physically realizable, namely the consistency condition, which states that any marginal distributions for observables that belong to different contexts are independent of a chosen measuring context:
\ben\label{consy}
\forall_{i,j} \sum_{\lambda_i \backslash \lambda_i \cap \lambda_j} p(\lambda_{i}) = \sum_{\lambda_j \backslash \lambda_i \cap \lambda_j} p(\lambda_{j}).
\een
The boxes that fulfill the consistency condition we call {\it consistent boxes}, and the set of all consistent boxes we denote as $\B$. Throughout the paper while referring to the set of boxes we mean exclusively the set of consistent boxes $\B$, without an explicit indication.

At this stage it should be noticed that the term box is more general than the classical probabilistic model for the whole set of observables $\mathcal{M}$, which is given by the joint probability distribution $p(\lambda|\mathcal{M})$, such that for all contexts
\ben
p(\lambda_c) = \sum_{\lambda \backslash \lambda_c} p(\lambda |\mathcal{M}).
\een
A set of valuable resources from this perspective are those boxes, which {\it cannot} be modeled classically, i.e., they cannot be described by a single joint probability distribution for all observables. The set of boxes that constitutes a set of resources we denote as $\B_v$, while its elements, i.e., particular valuable boxes, we will denote as $B_v$.
Otherwise, the set of classical boxes, which are useless from the point of view of valuable resources, we denote as $\B_{nv}$, whereas its elements by $B_{nv}$.
Furthermore, it is known that each box $B_{nv}\in \B_{nv}$ can be decomposed into a mixture of deterministic boxes which constitute the extremal points of the convex polytope that is identified with $\B_{nv}$. The extremal points of the set $\B_{nv}$ we will denote as $E_{nv}$, and each box from $\B_{nv}$ can be written as $B_{nv} = \sum_i p_i E^i_{nv}$ for a proper distribution $\{ p_i \}$. Notice that in general the extremal boxes $E_{nv}$ are not all the extremal vertices of the set of consistent boxes $\B$. The extremal boxes of $\B$ which do not belong to the set $\B_{nv}$, and constitutes valuable resources, we will denote as $E_{v}$.

Let us now formalize the resource theory approach regarding to the notion of contextuality. The crucial aspect of contextual systems is that one cannot ascribe the values for each observable into the physical system prior measurements such that they will obey the observed statistics. Otherwise, i.e., if one could in principle attribute the observed values of all measurements it would mean that the construction of a probabilistic model $p(\lambda|\mathcal{M})$ is possible. Therefore, if we regard a set of contextual boxes $\B_{C}$ as a set of resources, then the set of noncontextual boxes $\B_{NC}$ composes of valueless objects, $\B_{nv} \equiv \B_{NC}$.

Suppose now that we have instances of physical systems composed of two (or in general more than two) subsystems such that each subsystem is being measured by two spatially separated parties, Alice and Bob. Due to space-like separation, each pair of measurements performed by the two parties $x\in \M_A,y \in \M_B$ commute, hence gives naturally arising contexts $c=(x,y)$. From this perspective, the boxes are given by families of probability distributions $p(\lambda_c)\equiv p(a,b|x,y)$, where $\lambda_c = (a,b)$ is a pair of measurement outcomes of measurements $(x,y)$. The comprehensive resource theory of nonlocality was presented in Ref. \cite{Vin14}, where nonlocal boxes were identified as valuable resources, whereas local behaviors were considered as the boxes from the set $\B_{nv}$ (see also Ref. \cite{Fine}).

With respect to nonlocality, the consistency conditions \eqref{consy} are the strict analogues of nonsignaling conditions: the marginal distributions for one party are independent of the measurement choice made by spatially separated other party. Hence the consistency conditions assures nonsignalling in the case when measurement context arises from space-like separation of different parties and in this case the set of consistent boxes is to be understood simply as a set of nonsignaling boxes.



\subsection{Axioms of box-resource theories}

Once we wish to use the boxes for some specific tasks, we need to specify what are the permitted transformations by which we can process the boxes. Regarding the different sets of boxes, which either constitute valuable resources ($\B_{v}$), or not ($\B_{nv}$), we formulate two axioms that the operations must obey:

(O1) for {\it general operations} $\T$, given by normalization-preserving and consistency-preserving transformations $\T$ we must have
\ben
\forall_{B_1 \in \B_1} \quad \exists_{B_2 \in \B_2} \quad \T(B_1) = B_2.
\een
Notice that general transformations $\T$ transform boxes from one set into another without the assumption of the preservation of the dimensionality of the boxes. In particular some operations $\T$ may act as changing the number of contexts of a box by adding or removing some observables from the set $\M$.
In case when the dimensionality of boxes (the number of inputs with the respective number of outputs) is preserved, then the general operations can be described by a matrix form \cite{Barrett} (provided that the consistency of a transformed box is preserved):
\ben
\T \equiv \left( \begin{tabular}{c|c|c}
  $\alpha_{11}\T_{11}$ & $\cdots$ & $\alpha_{1|c|}\T_{1|c|}$ \\ \hline
  $\vdots$ & $\ddots$ & $\vdots$ \\ \hline
  $\alpha_{|c|1}\T_{|c|1}$ & $\cdots$ & $\alpha_{|c||c|}\T_{|c||c|}$ \\
\end{tabular} \right),
\een
where $\T_{ij}$ are stochastic matrices acting on the vector of a probability distribution $p(\lambda_j)$ from a box $B$, $0 \leq \alpha_{ij} \leq 1$, and $\sum_{j} \alpha_{ij} = 1$.

(O2) for {\it free operations} $\LL$, which constitutes a subset of general operations $\T$, we must have
\ben
\forall_{B_1 \in \B_{nv}} \quad \exists_{B_2 \in \B_{nv}} \quad \LL(B_1) = B_2.
\een
Similarly as a set of LOCC (local operations and classical communication) in the context of a resource theory of entanglement \cite{Hor4}, and a set of WCCPI (wirings and classical communication prior to inputs) in a resource theory of nonlocality \cite{Vin14,Allcock-wires}, the set of free operations $\LL$ is composed of those ``given for free'' operations in a device independent information processing, which by themselves cannot produce a resource from the set $\B_{nv}$. 

An important class of general operations are reversible operations $\R$, for which
\ben
\exists_{R, R^{-1} \in \R} \quad R^{-1}(R(B)) = B.
\een
A particular example of operations from the set $\R$ are relabelings (see \cite{Vin14} for details).

While considering boxes as resources we arrive at the question whether using general operations we can obtain resource that differs quantitatively from the original box. Similarly as in the entanglement theory we need to specify what are the relations between different resources, i.e., whether we can state quantitatively that a given resource is more valuable than the other.
It is therefore desirable to have a measure to quantify the value of the given resource.
Below we formulate the axioms that a reliable and usable measure for a given box, call it $X(B)$, need to fulfill.

First of all we need to know which boxes do not constitute a valuable resource. Thus the basic requirement for the measure $X$ is

(M1) {\it faithfulness}, which indicates
\ben
\forall_{B \in \B_{nv}} \quad X(B) = 0.
\een
Note that we restricted the axiom of faithfulness solely to the condition given above, and do not set an additional requirement $X(B_v) > 0$. After Ref. \cite{Vin14} we recall that some measure defined by usefulness with respect to a given operational task may give $X(B) = 0$ even for boxes which do not belong to the set $\B_{nv}$.

Another property that the measure need to have must reflect the fact that single-box general operations cannot increase the value of a given resource:

(M2) {\it monotonicity}, which states that
\ben
\forall_{B \in \B} \quad X(\T(B)) \leq X(B).
\een

As we have seen earlier, there is a set of reversible operations $\R$ such that each operation $R$ from this set has its inverse $R^{-1}$ for which there holds $R^{-1}(R(B))=B$. Since we already stated that a measure $X$ should be monotonic under general operations $\T$, the measure $X$ needs to fulfil

(M3) {\it partial invariance}, which means that the measure $X$ is invariant with respect to reversible operations $\R$ performed on a box
\ben
\forall_{R \in \R} \quad X(R(B)) = X(B).
\een

For the purpose of defining the next axiom we need to specify the distance measure for a pair of boxes.

{\definition The trace distance between two boxes $B$ and $B'$ is given by \cite{CT}:
\ben\label{trdist}
||B - B'|| := \sup_{\S} || \S(B) - \S(B')  ||_D,
\een
where $||.||_D$ denotes the {\it norm of difference between two probability distributions} which is given by variational distance between them
\be
||p_B(\lambda_c) - p_{B'}(\lambda_c)||_D: = \sum_{\lambda_c} |p_B(\lambda_c) - p_{B'}(\lambda_c)|,
\ee
where $p_{B}(\lambda_c)$ ($p_{B'}(\lambda_c)$) is the distribution of a measured context $c$ of a box $B$ ($B'$) for the outputs $\lambda_c$.
The supremum in \eqref{trdist} is taken over all operations which transform boxes into a probability distribution according to
\ben
\S(B):= \sum_{c} \alpha_{c} \T_{c} p_B(\lambda_c),
\een
where $\T_{c}$ are stochastic matrices, $0 \leq \alpha_{c} \leq 1$, and $\sum_{c} \alpha_{c} = 1$.
}

To this end we notice that any experimental realization of boxes involves inevitable distortion from the ideal box we want to realize. Suppose that an experimentally realized imperfect box $B'$ is close to the target box $B$, which constitutes a valuable resource:
\ben\label{bliskie}
||B - B'|| \leq \epsilon,
\een
where $||.||$ denotes the {\it trace distance between two boxes} \cite{CT}. Now, the desired property of the measure that conforms the requirements raised by the non-ideal experimental processes is:

(M4) {\it asymptotic continuity}, which means that for two close boxes \eqref{bliskie}, there holds
\ben
|X(B) - X(B') | \leq \epsilon \log d + f(\epsilon),
\een
where $d$ is dimensionality of the boxes and $f$ is a function such that $f(\epsilon) \xrightarrow{\epsilon \rightarrow 0} 0$.

Apart from the above axioms we recall also two welcome properties that the quantifiers of a resource should satisfy:

(P1) {\it convexity}, which means that for any box $B$ which is decomposable into $B = \sum_i p_i B_i$, there is
\ben
X(\sum_i p_i B_i) \leq \sum_i p_i X(B_i),
\een
where $B_i$ are arbitrary boxes.

If we consider a measure to be extensive we would also require the property termed by

(P2) {\it additivity}, for which a measure fulfills
\ben
\forall_{B_1 \in \B_1,B_2 \in \B_2} \quad X(B_1 \otimes B_2) = X(B_1) + X(B_2),
\een
where $\otimes$ denotes tensor product of boxes, i.e., for boxes $B_1=\{ p(\lambda_c)\}$ and $B_2=\{ q(\lambda_{c'}) \}$ we have $B_1 \otimes B_2 = \{ p(\lambda_c)q(\lambda_{c'}) \}$.

\section{Measures of contextuality}

In this section we focus on contextuality measures as defined in Ref. \cite{context-measures}. We then introduce the tools needed for proving the asymptotic continuity of the measures.
The crucial assumption for the latter is closeness of two respective boxes. As we shall see the assumption is still valid while considering two close quantum states from which the respective boxes are drawn.

Given a single joint probability distribution of an arbitrary context $c$ of a box $B$ we can trivially define an extended joint probability distribution for all observables in $\mathcal{M}$:
\ben
\mathcal{A}_B(c):=p(\lambda_c)P(\lambda_r|c),
\een
where $P(\lambda_r|c)$ is a joint probability distribution for all the observables from the set $\mathcal{M}\backslash c$, and $\lambda_r$ is the corresponding set of measurement outcomes.

{\definition For a given box $B= \{p_{B}(\lambda_c)\}$ we call its extension a family of distributions:
\be
\F(B):=\{\mathcal{A}_B(c)\},
\label{eq:ExtB}
\ee
where $\A_{B}(c):=p_{B}(\lambda_c)P(\lambda_r|c)$ is an extension of distribution $p_{B}(\lambda_c)$ to all observables of a box $B$.
}

For further purposes we will write $\E_{p(c)}(B)$ to denote a distribution, related to the extension $\F(B)$, which for the ease of notation we model as a quantum state:
\ben
\E_{p(c)}(B):=\sum_c p(c) |c\>\<c|\ot \mathcal{A}_{B}(c),
\een
for an arbitrary probability distribution $p(c)$ of contexts.

We will now recall the definitions of two measures of contextuality as presented in \cite{context-measures}. The first measure---mutual information of contextuality---captures the idea that a contextual system cannot be described by a single joint probability distribution for all observables that can be measured on the system. It quantifies the correlations between the different joint probability distributions consistent with each of the measured context and the number of a chosen context. The second measure---relative entropy of contextuality---is defined in terms of a statistical distance between a set of probability distributions describing a contextual system to the closest single noncontextual joint probability distribution. The relative entropy of contextuality is a natural extension to contextual systems an analogous measure of nonlocality presented in \cite{DGG}, called \textit{statistical strength of nonlocality proofs}.

{\definition
Mutual information of contextuality of a given box $B$ we call
\ben\label{mic}
I_{\max}(B) := \sup_{\{p(c) \}} \inf_{\{\mathcal{A}_B(c) \}} I (\sum_c p(c) |c\>\<c| \otimes \mathcal{A}_B(c) ),
\een
where $I$ is mutual information between probability distributions.
}

We also have
{\definition
Relative entropy of contextuality of a given box $B$ we call
\ben\label{rec}
X_{\max}(B) := \sup_{\{p(c) \}} \inf_{\{p(\lambda) \}} \sum_c p(c) D \left( p_B(\lambda_c) || p(\lambda_c)   \right),
\een
where $D$ is relative entropy, and infimum is taken over all joint probability distributions for all observables in $\mathcal{M}$, $p(\lambda)$, such that $p(\lambda_c)$ is a proper marginal distribution for a given context:
\ben
p(\lambda_c) = \sum_{\mathcal{M}\backslash c} p(\lambda).
\een
}

Note that given a box $B$, we have:
\ben
I_{\textrm{max}}(B) = X_{\textrm{max}}(B),
\een
hence we can use $I_{\textrm{max}}(B)$ and $X_{\textrm{max}}(B)$ interchangeably.
This equality will prove useful, as the arguments given in \cite{Synak-Horodecki} that REC is asymptotic continuous are not sufficient for a proof of the latter fact. Instead will prove the asymptotic continuity of MIC.

A specific measure of contextuality also utilizing the concept of relative entropy is given in the following.
{\definition
Uniform relative entropy of contextuality of a given box $B$ we call
\ben\label{recu}
X_{\uu}(B) :=  \inf_{\{p(\lambda) \}} \sum_c \frac1n D \left( p_B(\lambda_c) || p(\lambda_c)   \right),
\een
where $n$ is a number of different contexts of a box $B$.
}

We now observe the following fact, namely when the trace distance of two boxes is small then for any extension of one box there exists an extension of the other, which is close to the former.

{\observation If two boxes $B=\{p_{B}(\lambda_c)\}$ and $B'=\{p_{B'}(\lambda_c)\}$ satisfy
$||B - B'|| \leq \delta$,
then for $\delta> 0$ and for any fixed $\{p(c)\}$ we have
\be
\forall_{\E_{p(c)}(B)}\quad \exists_{\E_{p(c)}(B')} \quad||\E_{p(c)}(B)-\E_{p(c)}(B')||_D \leq \delta.
\label{eq:ExtB1ExtB2}
\ee
\label{obs:ExtB1ExtB2}
}

{\it Proof.}
The left-hand side of ineq. (\ref{eq:ExtB1ExtB2}) equals:
\be
\sum_c p(c) ||\mathcal{A}_{B}(c) - \mathcal{A}_{B'}(c)||_D,
\label{eq:ex-1}
\ee
where
\ben
\mathcal{A}_{B}(c)&:=& p_{B}(\lambda_c)P(\lambda_r|c), \\
\mathcal{A}_{B'}(c)&:=& p_{B'}(\lambda_c)P(\lambda_r|c). \label{Aprim}
\een
Consider the difference between the two distributions $\mathcal{A}_{B}(c)$ and $\mathcal{A}_{B'}(c)$:
\ben
\lefteqn{||\mathcal{A}_{B}(c) - \mathcal{A}_{B'}(c)||_D} \nonumber \\
&=& \sum_{\lambda_c,\lambda_r} | p_{B}(\lambda_c)P(\lambda_r|c) - p_{B'}(\lambda_c)P(\lambda_r|c)| \nonumber \\
&=& \sum_{\lambda_c} ( |p_{B}(\lambda_c)-p_{B'}(\lambda_c)|(\sum_{\lambda_r} P(\lambda_r|c))) \nonumber \\
&=& \sum_{\lambda_c}  |p_{B}(\lambda_c)-p_{B'}(\lambda_c)| \nonumber \\
&=& || \S^*(B) - \S^*(B') ||_D \nonumber \\
&\leq& \sup_{\S}|| \S(B) - \S(B') ||_D \nonumber \\
&\leq& \delta,
\label{eq:ex-2}
\een
where the first inequality comes from the fact that $P(\lambda_r|c)$ is a probability distribution for each $c$, $\S^*$ is such that $\alpha_{c}=1$ and $\T_{c}$ is the identity matrix for a chosen context $c$, while the last inequality follows from the assumption of the observation. Using the last inequality back in (\ref{eq:ex-1}) we obtain (\ref{eq:ExtB1ExtB2}), which ends the proof.\blacksquare

We need to introduce the notation for the corresponding extensions given by eq. \eqref{Aprim}, as it is described below.

{\definition Consider two boxes $B=\{p_{B}(\lambda_c)\}$ and $B'=\{p_{B'}(\lambda_c)\}$. For any extension $\F(B)$ the extension $\F(B')$ given by Eq. \eqref{Aprim}, which satisfy $||\E_{p(c)}(B)-\E_{p(c)}(B')|| \leq \delta$, we will denote as $\F(B'|B)$.}

In the next section we will derive asymptotic continuity of mutual information of contextuality $I_{\textrm{max}}$. As we shall see in the following, the closeness of quantum states, $\rho$ and $\rho'$, implies the closeness of the respective boxes, $B$ and $B'$, while assuming perfect measurements on quantum states. Consider two boxes, $B$ and $B'$, each of them drawn by the same set of measurements $M_c$, but on a different quantum state $\rho$ and $\rho'$, respectively, such that $||\rho- \rho'|| \leq \delta$, where in case of matrices $||A||:=\sqrt{AA^{\dag}}$. Then $I_{\textrm{max}}$ can be proved asymptotic continuous with respect to trace distance for quantum states $||\rho - \rho'|| < \delta$.

{\observation Consider any two states $\rho,\rho'$ such that $||\rho - \rho'|| <\delta$.
Let ${\cal O}=\{M_c\}$ be a set of operators, which generates two respective boxes $B$ and $B'$ on states $\rho$ and $\rho'$, respectively,
in such a way that each distribution of the box $B$ ($B'$) is given by $p_B(\lambda_c)\equiv \{\textrm{Tr}M_c \rho \}$ ($p_{B'}(\lambda_c)\equiv \{\textrm{Tr}M_c \rho' \}$).
Then we have:
\ben
||B - B' || < 2\delta.
\een
\label{obs:box-state}
}

{\it Proof.}
Consider a distribution $\alpha^*_c$ and stochastic matrices $\T^*_c$ that realizes supremum in definition of $||B - B' ||$. Let $M^*$ be a measurement operator defined as
\ben
M^*:=\sum_c \alpha^*_c \T^*_c M_c.
\een
From the assumption we have that
\ben
\delta &>& ||\rho  - \rho'|| \nonumber \\
&=& \sup_{\tilde{M}} \textrm{Tr} \tilde{M}(\rho  - \rho') \nonumber \\
&\geq& \textrm{Tr} P^+ M^*(\rho  - \rho'),
\een
where $P^+$ is a projector onto the positive subspace of $M^*(\rho  - \rho')$. The RHS of the last inequality above is equal to
\ben
S^+ \equiv \sum_{\{+\}}(r_{\rho}-r_{\rho'}),
\label{eq:Splus}
\een
where $r_{\rho} = Tr M^* \rho$ ($r_{\rho'} = Tr M^*\rho'$) and the sum is over all terms $r_{\rho}-r_{\rho'}>0$.
Changing the roles of $\rho$ and $\rho'$ we obtain analogously:
\ben
\delta &>& ||\rho'  - \rho|| \nonumber \\
&=& \sup_{\tilde{M}} \textrm{Tr} \tilde{M}(\rho'  - \rho) \nonumber \\
&\geq& \sum_{\{+\}}(r_{\rho'}-r_{\rho}) \nonumber \\
&=& |\sum_{\{-\}}(r_{\rho}-r_{\rho'})| \nonumber \\
&\equiv& S^-.
\een
From the above inequality, and (\ref{eq:Splus}), we have that $\max (S^+, S^-)<\delta$, and since the distributions $r_{\rho}$ and $r_{\rho'}$ are obtained with the optimal $\alpha^*_c$ and $\T^*_c$, we obtain
\begin{equation}
||B - B'|| = ||r_{\rho} - r_{\rho'}||_D = S^+ + S^- \leq 2\max (S^+, S^-) < 2 \delta,
\end{equation}
which ends the proof.\blacksquare

\section{Asymptotic continuity of $I_{\max}$}

In this section we will prove the asymptotic continuity for the measure of contextuality given by mutual information of contextuality \eqref{mic}. We will focus on MIC rather than \textit{relative entropy of contextuality}, and we will introduce a novel method of proving the asymptotic continuity. Nevertheless, the asymptotic continuity of the former (MIC) implies the same for the latter (REC) since the two measures are equal to one another \cite{context-measures}.

At this stage let us also recall the property of asymptotic continuity for von Neumann entropy. Consider two quantum states $\rho_1$ and $\rho_2$ of dimension $d$, for which
\ben
||\rho_1 -\rho_2||\leq 1/2,
\een
where $||.||$ is trace norm.
Von Neumann entropy is asymptotically continuous since
\be
|S(\rho_1) -S(\rho_2)|\leq ||\rho_1 -\rho_2|| \log d + \eta(||\rho_1-\rho_2||),
\label{eq:entropy}
\ee
where $\eta(x)=-x\log x$.

Furthermore, quantum conditional entropy is also asymptotically continuous \cite{Alicki-Fannes}, i.e., for $||\rho_1 -\rho_2||\leq \epsilon$ we have
\ben
\lefteqn{|S_{X|Y}(\rho_1) -S_{X|Y}(\rho_2)|\leq} \\
&& 4||\rho_1 -\rho_2|| \log d + 2\eta(||\rho_1-\rho_2||) + 2\eta(1-||\rho_1-\rho_2||), \nonumber
\label{eq:conditionalentropy}
\een
where $X,Y$ denotes two subsystems of the states $\rho_1$ and $\rho_2$.

Now, Observation \ref{obs:ExtB1ExtB2} allows us to write the following.

{\observation If for some $\delta> 0$ two boxes $B=\{p_{B}(\lambda_c)\}$ and $B'=\{p_{B'}(\lambda_c)\}$ satisfy
\be
||B - B'|| \leq \delta,
\ee
then for any fixed $\{p(c)\}$ we have
\be
\forall_{\E_{p(c)}(B)}\quad \exists_{\E_{p(c)}(B'| B)} \quad|I(\E_{p(c)}(B))-I(\E_{p(c)}(B'| B))| \leq g(\delta),
\label{eq:IExtB1IExtB2}
\ee
with
\be
g(\delta)=5\delta \log d + 2 \eta(1-\delta)+3\eta(\delta),
\label{eq:IB-ascont}
\ee
where $I$ is mutual information, $d=\min(\prod_{i=1}^{k}a_i,|E_G|)$, where $\prod_{i=1}^{k}a_i$ and $|E_G|$ are the dimensions of the two subsystems of the box $B$, respectively.
\label{obs:IExtB1IExtB2}
}

{\it Proof.}
Due to Observation \ref{obs:ExtB1ExtB2} for every $\E_{p(c)}(B)$ there exists $\E_{p(c)}(B'| B)$ such that $||\E_{p(c)}(B)-\E_{p(c)}(B'| B)|| \leq \delta$.
Using the definition of mutual information, the left-hand side of (\ref{eq:IExtB1IExtB2}) can be written as
\ben
\lefteqn{|I(\E_{p(c)}(B))-I(\E_{p(c)}(B'|B))|} \nonumber \\
&&=|S_X(\E_{p(c)}(B))-S_{X|Y}(\E_{p(c)}(B)) + \nonumber \\
&&-S_X(\E_{p(c)}(B'| B))+S_{X|Y}(\E_{p(c)}(B'| B))|.
\een
Then, making use of a triangle inequality we can bound this expression by
\ben
\lefteqn{|S_X(\E_{p(c)}(B))-S_X(\E_{p(c)}(B'| B))|+} \nonumber \\
&+& |S_{X|Y}(\E_{p(c)}(B))-S_{X|Y}(\E_{p(c)}(B'| B))| \nonumber \\
&=& 5\delta \log d + 2 \eta(1-\delta)+3\eta(\delta),
\een
where in the last step we used asymptotic continuity of von Neumann entropy (\ref{eq:entropy}) as well as quantum conditional entropy (\ref{eq:conditionalentropy}). \blacksquare



Consider a following lemma:

{\lemma For any real valued function $f$, and for any two sets $T$ and $T'$, if there exists a real valued function $g$, such that for any positive $\delta$ and any $\rho \in T$ there exists $\sigma_\rho \in T'$, such that:
\be
|f(\rho) - f(\sigma_{\rho})|\leq g(\delta),
\label{eq:assumption1}
\ee
and for any $\sigma \in T'$ there exists $\rho_{\sigma}\in T$, such that:
\be
|f(\rho_{\sigma}) - f(\sigma)|\leq g(\delta),
\label{eq:assumption2}
\ee
then there holds:
\be
|\inf_{\rho \in T} f(\rho) - \inf_{\sigma \in T'} f(\sigma)|\leq g(\delta) + \delta,
\ee
and
\be
|\sup_{\rho \in T} f(\rho) - \sup_{\sigma \in T'} f(\sigma)|\leq g(\delta) + \delta,
\ee
provided that $\inf_{\rho \in T} f(\rho)$ and $\inf_{\rho \in T'} f(\sigma)$, as well as
$\sup_{\rho \in T} f(\rho)$ and $\sup_{\rho \in T'} f(\sigma)$ are bounded.
\label{lem:inf-ascont}
}

We prove the lemma in Appendix \ref{APP1}.
We can now state the following theorem:

{\theorem For any fixed $p(c)$ the function
\ben
I_{p(c)}(B) = \inf_{\{\mathcal{A}_{B}(c)\}} I(\sum_c p(c)|c\>\<c|\ot \mathcal{A}_{B}(c))
\een
is asymptotically continuous, i.e.,
for any $||B - B'||\leq \delta$ there is
\ben
|I_{p(c)}(B) - I_{p(c)}(B')|\leq g(\delta) + \delta,
\een
with $g(\delta)$ given in the right-hand side of (\ref{eq:IB-ascont}).
\label{thm:ipc-as-cont}
}

{\it Proof.}
Assume that $||B - B'||\leq \delta$ for two boxes $B = \{g_{B}(c)\}$ and $B' = \{g_{B'}(c)\}$. Let us first consider $I_{p(c)}(B)$. By definition of infimum, there exist a sequence of extensions of distributions $\F^n(B)=\{A_{B}^n(c)\}$, such that
\ben
\lim_{n\rightarrow \infty} I(\mathcal{E}^n(B)) = I_{p(c)}(B),
\een
and similarly there exist $\F^n(B')=\{A_{B'}^n(c)\}$, such that
\ben
\lim_{n\rightarrow \infty} I(\E^n_{p(c)}(B')) = I_{p(c)}(B').
\een

Let us now specify the following two sets:
\ben
T &:=& \bigcup_{n=1}^{\infty}\{ I(\E^n_{p(c)}(B)),I(\E^n_{p(c)}(B| B'))\}, \\
T' &:=& \bigcup_{n=1}^{\infty}\{ I(\E^n_{p(c)}(B'| B)),I(\E^n_{p(c)}(B'))\},
\een
and also denote $V$ ($V'$) as the set of $I(\E_{p(c)}(B))$ ($I(\E_{p(c)}(B'))$) for \textit{all} distributions $\mathcal{A}_{B}(c)$ ($\mathcal{A}_{B'}(c)$).
Notice that $\{I(\E^n_{p(c)}(B))\} \subset T$, and therefore we have that
\ben\label{in1}
\inf T \leq \inf \{I(\E^n_{p(c)}(B))\} = I_{p(c)}(B).
\een
On the other hand, $T \subset V$, and therefore
\ben\label{in2}
I_{p(c)}(B) = \inf V \leq \inf T,
\een
where the equality is by definition of $I_{p(c)}(B)$. The inequalities \eqref{in1} and \eqref{in2} leads to
\ben
\inf T = I_{p(c)}(B),
\een
and the same reasoning also gives
\ben
\inf T' = I_{p(c)}(B').
\een

Now, due to Observation \ref{obs:IExtB1IExtB2}, there is
\ben\label{eq:closetoo}
|I(\E^n_{p(c)}(B)) - I(\E^n_{p(c)}(B'| B))|\leq g(\delta), \\
|I(\E^n_{p(c)}(B')) - I(\E^n_{p(c)}(B| B'))|\leq g(\delta).
\label{eq:closeto}
\een
Let us take in the assumption of Lemma \ref{lem:inf-ascont} the function $f: \mathbf{R}\rightarrow \mathbf{R}$ to be simply the identity function $f(x)=x$ (it is easy to see that other assumptions are satisfied by the construction and the fact that the boxes $B$ and $B'$ are close to each other and by inequalities \eqref{eq:closetoo}-\eqref{eq:closeto}).
Then, by Lemma \ref{lem:inf-ascont} we obtain
\be
|\inf_{t\in T} t - \inf_{t'\in T'} t'|\leq g(\delta) + \delta,
\ee
which is exactly what we want to prove.\blacksquare

We can now state the main theorem of this section. The result is crucial from the experimental point of view. Namely it states that for two close boxes (one of which is the ideal box we aim to obtain, while the second is its imperfect experimental realization) the measure of the amount of contextuality of the experimentally obtained box cannot differ from the measure of contextuality of the target box by more than the distance of the two boxes with a factor depending only logarithmically on the dimension of them.

{\theorem The measure of contextuality $I_{\max}$ is asymptotically continuous, i.e., for any two boxes $B$ and $B'$ fulfilling $||B - B'||\leq \delta$, there is
\ben
|I_{\max}(B) - I_{\max}(B')|\leq 6g(\delta) + 3\delta,
\een
with
\ben
g(\delta)=5\delta \log d + 2 \eta(1-\delta)+3\eta(\delta).
\een
\label{thm:main-as-cont}
}

{\it Proof.}
The proof follows the similar lines as the proof of Theorem \ref{thm:ipc-as-cont}.
Let us first consider $I_{\max}(B)$. By definition of supremum, there exist a sequence of distributions $\{p_n(c)\}$, such that
\ben
\lim_{n\rightarrow \infty} I_{p_n(c)}(B) = I_{\max}(B),
\een
and similarly there exist $\{p'_n(c)\}$, such that
\ben
\lim_{n\rightarrow \infty} I_{p'_n(c)}(B) = I_{\max}(B').
\een

Let us now specify the following two sets:
\ben
T &:=& \bigcup_{n=1}^{\infty}\{ I_{p_n(c)}(B),I_{p'_n(c)}(B)\}, \\
T' &:=& \bigcup_{n=1}^{\infty}\{  I_{p_n(c)}(B'),I_{p'_n(c)}(B')\},
\een
and also denote $V$ ($V'$) as the set of $I_{p_n(c)}(B)$ ($I_{p'_n(c)}(B')$) for \textit{all} distributions $\{p_n(c)\}$ ($\{p'_n(c)\}$).
Applying the similar reasoning as in the proof of Theorem \ref{thm:ipc-as-cont}, but this time concerning the suprema, we arrive at
\ben
\sup T = I_{\max}(B),
\een
and the same reasoning gives also
\ben
\sup T' = I_{\max}(B').
\een

Now, due to Theorem \ref{thm:ipc-as-cont}, there is
\ben\label{eq:closetoss}
|I_{p_n(c)}(B) - I_{p_n(c)}(B')|\leq g(\delta) +\delta, \\
|I_{p'_n(c)}(B') - I_{p'_n(c)}(B)|\leq g(\delta) + \delta.
\label{eq:closetos}
\een
Let us take in the assumption of Lemma \ref{lem:inf-ascont} the function $f: \mathbf{R}\rightarrow \mathbf{R}$ to be simply the identity function $f(x)=x$ (it is easy to see that other assumptions are satisfied by the construction and the fact that the boxes $B$ and $B'$ are close to each other and by inequalities \eqref{eq:closetoss}-\eqref{eq:closetos}).
Then, by Lemma \ref{lem:inf-ascont} we obtain
\be
|\sup_{t\in T} t - \sup_{t'\in T'} t'|\leq g(\delta) + 2\delta,
\ee
which is exactly what we want to prove.\blacksquare

From the two main theorems presented here, we have immediate corollary, which states, that the relative entropy of contextuality is
continuous with respect to quantum states, provided ideal measurements.

{\corollary For two quantum states $\rho$ and $\sigma$ such that $||\rho - \sigma|| \leq \delta$, for any set of quantum measurements ${\cal M}=\{M_c\}$ which generates on the states respective boxes $B =\{Tr M_c \rho\}$ and $B' =\{Tr M_c \sigma\}$, we have:
\ben
|I_{p(c)}(B) - I_{p(c)}(B')|\leq 2g(\delta) + 2\delta,
\een
as well as
\ben
|I_{\max}(B) - I_{\max}(B')|\leq 12g(\delta) + 6\delta,
\een
with $g(\delta)=5\delta \log d + 2 \eta(1-\delta)+3\eta(\delta)$.
}

{\it Proof}. This corollary follows directly from observation \ref{obs:box-state}, and theorems \ref{thm:ipc-as-cont} and \ref{thm:main-as-cont}.

\section{Computable upper bound on the measure of a resource}

In this section we will connect two measures of box resources: the one which is based on relative entropy, with the other which reports how much costs the creation of a box. In case of the resource which is contextuality, we will show, that the relative entropy of contextuality
for chain boxes is upper bounded (up to a normalization factor) by the cost of contextuality. In derivation of this result we will use the property of (P1) convexity of the relative entropy of contextuality, which was observed in \cite{PR}, but without a formal proof, which we provide in Appendix \ref{app:conv}.

To achieve this, we will show a general result, which holds for a measure satisfying certain axioms. First, we assume that sets $\B$ and $\B_{nv}$ are convex
polytopes and $\B_{nv} \subset \B$. We will now need the notion of a {\it convex roof of a measure} which for the measure $X$ is defined as
\be
\hat{X}(B): = \inf_{\{p_i, E_i\}} \sum_ i p_i X(E_i),
\ee
where $E_i$ are extremal boxes of the polytope of all consistent boxes, and infimum is taken over all ensembles
of the box $B$ with all extremal boxes, so that $ \sum_i p_i E_i = B$.

We begin with noticing the following observation:

{\observation For any convex measure $X$ and $B \in \B$ we have
\be
X(B) \leq {\hat X}(B).
\ee
\label{obs:conv-roof}
}

{\it Proof.} It follows from the definition of $\hat{X}$.\blacksquare

Next we note that some of the extremal boxes of the polytope of consistent boxes are valuable ($E^i_v$), while the others are not ($E^i_{nv}$).
Such a polytope $\B$ in some cases can satisfy the following property, which we call {\it vertex-equivalence property}:
\be
\forall_{E_v,E'_v \in \B} \quad \exists_{R\in{\cal R}} \quad R(E_v) = E_v',
\ee
where $E_v,E_v'$ are extremal, valuable boxes. In other words, any valuable box can be transformed into any other valuable box by means of reversible operations.

We will now show that if the polytope $\B$ has the vertex-equivalence property, a convex measure $X$ satisfying axioms (M1) and (M3),
is upper-bounded (up to a normalization factor $X(E_v)$ for some extremal valuable box $E_v$) by another measure which we will call the {\it cost of the resource}. We first formalize the latter measure, as a generalization of the well known measure of the {\it cost of non-locality}:

{\definition For a box $B \in \B$ the cost of the resource for this box
is
\be
C(B) = \inf_p \{p: B = pB_v + (1-p) B_{nv},\, B_v \in \B,\, B_{nv} \in \B_{nv}\}.
\ee
}

We are ready to present the main result of this section:

{\proposition Let $\B$ be a polytope of consistent boxes satisfying the vertex equivalence property. Also let $X$ be a measure of resource acting on boxes $B \in \B$, which satisfies the axioms of (M1) faithfulness, (M3) local invariance and (P1) convexity. We have then
\be
X(B) \leq C(B) X(E_v),
\ee
where $E_v$ is arbitrarily fixed, extremal, valuable box in $\B$, and $C(.)$ is the cost of the resource.
\label{prop:main-bound}
}

{\it Proof.}
Let us fix $B \in \B$ arbitrarily. Since $X$ is convex, by Observation \ref{obs:conv-roof}, we have:
\be
X(B) \leq \inf_{\{p_i, E_i\}}  \sum_i p_i X(E_i).
\ee
Now, for all $E_i \in \B_{nv}$ we have by axiom (M1) $X(E_i) = 0$. Thus we have
\be
X(B) \leq \inf_{\{p_i, E_i\}}  \sum_{i\in \I} p_i X(E_i),
\ee
where $\cal I$ is the set of indices for an ensemble $\{ p_i, E_i\}$ such that if $i\in {\cal I}$ then $E_i$ is valuable box.
On the other hand, by the vertex equivalence property of the polytope $\B$ there is $X(E_i) = X(E_v)$ for all $i \in {\cal I}$
and $E_v$ being some extremal valuable box fixed arbitrarily. Indeed, all the valuable boxes have the same value of the
measure $X$, since they are reversibly transformable one into another, and the measure $X$ satisfies the axiom (M3), i.e.,
does not change under such transformations. Hence
\be
X(B) \leq X(E_v) \inf_{\{p_i, E_i\}}  \sum_{i\in \I} p_i,
\ee
and furthermore
\be
\hat{X}(B) = X(E_v) \inf_{\{p_i, E_i\}}  \sum_{i\in \I} p_i.
\label{eq:conv-bound}
\ee
We will now prove that the minimal $\sum_{i \in \I} p_i $ equals exactly $C(B)$, which makes the thesis.
Consider again any pure ensemble into boxes $E_i$ with probabilities $p_i$. We can perform a valid decomposition of a box $B$ into:
\be
B = p B_v + (1-p)B_{nv},
\ee
where $p= \sum_{i \in \I} p_i$ and, by definition of $\I$: $B_{v} = \sum_{i \in \I} p_i E_i/(\sum_{i \in \I} p_i)$, and $B_{nv} = \sum_{i \notin \I} p_i E_i/(\sum_{i \notin \I} p_i)$.
Thus, there is $C(B) \leq p$, and since the ensemble was arbitrary, we have that
\be
C(B) \leq \inf_{\{p_i, B_i\}} \sum_{i \in \I} p_i.
\ee
This, via (\ref{eq:conv-bound}), implies that
\be
C(B)X(B_v) \leq \hat{X}(B).
\label{eq:one-part}
\ee

We now prove converse inequality.
To begin with, consider any decomposition which achieves $C(B)$ (the case when infimum in its definition is not attained, is then obvious):
\be
B= q B_{v} + (1-q)B_{nv}.
\ee
Now, consider a decomposition of a box $B_v$ into extremal boxes. Since $q = C(B)$, there cannot be any deterministic box
in the decomposition of $B_v$, or else we would have smaller value of $C(B)$ by subtracting this box from $B_v$ and thereby lowering $q$.
Thus, we can write $B_v = \sum_j r_j E_j$, with $E_j$ - extremal valuable boxes.
Let us also decompose a box $B_{nv}$ into extremal non-valuable boxes (it is possible, since this box is by definition not valuable, and the set $\B_{nv}$ is a convex polytope by assumption): $B_{nv} = \sum_{k=1}^K s_k D_k$.
We then have an ensemble of the box $B$ of the form $\chi = \{ (qr_1,...,qr_N, (1-q)s_1,...,(1-q)s_K), E_1,..,E_N, D_1,..,D_K)\}$
for some natural numbers $N,K \geq 1$, i.e., by construction:
\be
B = q\sum_{j=1}^N r_j E_j + (1-q)\sum_{k=1}^{K} D_k.
\ee
Now, we have $X(D_k) =0$ for $k\in\{1,...,K\}$.
Consider then the set of indices $\I=\{1,...,N\}$.
Since $\hat{X}$ is defined as a function which is minimized over all ensembles of the box $B$, it is upper bounded by the value of the function
on the ensemble $\chi$. Thus we have:
\ben
\hat{X}(B) &=& \inf_{\{p_i,B_i\}} \sum_i p_i X(B_i) \nonumber \\
         &\leq& \sum_{i \in {\cal I}}q r_i X(E_i) \nonumber \\
            &=&  X(B_v)  \sum_{i}q r_i  \nonumber \\
            &=& X(B_v)  q, \label{eq:sequence}
\een
where the pre-last equality is due to the fact that for all $i \in {\cal I}$ $X(E_i)= X(B_v)$, since the polytope of consistent boxes satisfies the vertex equivalence property.
This proves that
\ben
\hat{X}(B) \leq C(B)X(B_v),
\label{eq:last}
\een
which together with the opposite inequality (\ref{eq:one-part}) proves the thesis.
\blacksquare

Let us note here that in case when $\B_{nv}$ is a polytope, the cost of the resource can be computed by linear programming, hence the above
proposition provides a computable upper bound on a convex measure satisfying (M1) and (M3).

In next section, we present two examples of polytopes, which satisfy the vertex-equivalence property, for which the above proposition applies.
We present now more rough bound, which holds for all polytopes, including those which do not satisfy the latter property.

{\proposition Let $\B$ be a polytope of consistent boxes. Also let $X$ be a measure of a resource acting on boxes $B \in \B$, which satisfies axioms of (M1) faithfulness and (P1) convexity. We then have
\ben
X(B) \leq C(B) \max_{E_v\in \B} X(E_v),
\een
where maximum is taken over all extremal, valuable boxes $E_v$ from $\B$, and $C(.)$ is the cost of the resource.
\label{prop:main-bound2}
}

{\it Proof.}
The proof goes along similar lines as the second part of the proof of Proposition \ref{prop:main-bound2}, namely starting right after the inequality (\ref{eq:one-part}). We then repeat the proof until the sequence of (in)equalities (\ref{eq:sequence}) and modify them while observing that for $s = \sum_{i\in \I} qr_i$
\be
\sum_{i\in \I} q r_i X(E_i) \leq \max_{i \in \I} X(E_i) \times s \leq \max_{E_v\in \B} X(E_v) \times s,
\ee
where in the first inequality maximum is taken over extremal valuable boxes $E_i$ from the particular ensemble of the box $B$, while in the second over all extremal valuable boxes from the polytope $\B$. According to the remaining part of the proof of Proposition \ref{prop:main-bound2}, we obtain:
\ben
\hat{X}(B) \leq C(B) \times \max_{E_v \in \B} X(E_v),
\een
which is less tight version of the inequality (\ref{eq:last}) as required.\blacksquare

\subsection{Examples of the upper bounds on the $X_{\max}$ via contextuality (non-locality) cost}

We can now apply the Proposition \ref{prop:main-bound} in two cases. The first is the case of
the set $\B$ equal to the set all bipartite non-signaling boxes with two binary inputs and two binary outputs (we will denote it
$\B(2\times 2)$). In this case,
it is known that the only extremal valuable boxes of $\B$ have a form:
\be B_{rst}(a,b|x,y) = \left\{ \begin{array}{ll}
         1/2 & \mbox{if $a\oplus b = xy\oplus rx \oplus sy \oplus t$} \\
        0 & \mbox{else}\end{array} \right.
\ee
where $a,b,x,y,r,s,t$ are binary.

It is then clear from the above form that all extremal valuable boxes can be transformed reversibly into the $PR$-box. Moreover the set of all
non valuable boxes, is called the set of {\it local} boxes, and it is a convex polytope $\B_{nv}(2\times 2) \subset \B(2\times 2)$. We have then the following corollary:

{\corollary\label{ce4} For any box $B\in \B(2\times 2)$ there holds:
\be
X_{\max}(B) \leq \inf_{\{p_i, B_i\}} \sum_i p_i X_{\max}(B_i) =  C(B)\times \log {4\over 3 },
\ee
where $B_i$ are extremal boxes and $C(B)$ is the cost of non-locality.
}

{\it Proof.}
It is straightforward to check that $X_{\max}$ satisfies the axioms (M1) and (M3), and, as it was mentioned in \cite{PR} and is proved in
Appendix \ref{app:conv}, $X_{\max}$ is convex. Moreover, as we have mentioned before, all the non-local boxes with two binary inputs and outputs can be transformed by local reversible operations into the $PR$-box, i.e., $B_{000}$, for which $X_{\max}(B_{000})= \log {4 \over 3}$ \cite{context-measures}. Hence, by Proposition \ref{prop:main-bound} we have the thesis.\blacksquare

We now generalize the above result to the case of resources of contextuality. We consider a class of boxes corresponding to $n$-cycle hypergraph (a class of $CH^{(n)}$ boxes) scenarios as presented in Ref. \cite{BC90,Weh,AQBCC13,context-measures}.
Any box $B \in CH^{(n)}$ is given by $B= \{ p_B(\lambda_c) \}$, where a probability distribution for each context $c$ can be written as
\ben
p_B(\lambda_{c_i}) &\equiv& p(m_i m_{i+1} |M_i M_{i+1}) \nonumber \\
 &=& \frac12 (1 + \<m_i m_{i+1} \>),
\label{eq:chn-description}
\een
with $m_j = \pm 1$, where we use the convention $m_n m_{n+1}= m_1 m_n$.

The corresponding polytope of boxes compatible with this hypergraph
will be called $\B^n$. Let us note here, that $\B^4 \equiv \B(2\times 2)$. It is known \cite{context-measures} that the contextuality measure of any extremal valuable box $E_v$ for arbitrary $n$ is given by
\ben
X_{\max}(E_v) = \log \frac{n}{n-1}.
\een

{\corollary For a box $B \in \B^n$ we have
\be
X_{\max}(B) \leq {\hat{X}_{\max}}(B) = C(B)\log \frac{n}{n-1}.
\label{eq:xmax-chain-bound}
\ee
}

{\it Proof.}
According to the description of a box $B \in CH^{(n)}$ given in eq. (\ref{eq:chn-description}), it can be uniquely described by a collection of $n$ correlators
\ben\label{vect}
B \equiv (\<m_1 m_{2} \>, \<m_2 m_{3} \>, ..., \<m_n m_{1} \>  ).
\een
Now, for any given $n$ we have $2^{n-1}$ extremal contextual boxes of the form \eqref{vect}, where $\<m_i m_{i+1} \> = \pm 1$, such that the number of negative components is odd \cite{AQBCC13}.
It now suffices to observe, that we can obtain all the others extremal contextual boxes from, e.g., $(-1,1,1,1,...,1,1)$ simply by bit-flipping the chosen outputs, which is a contextuality preserving operation (a particular form of relabeling the outputs). Indeed, by performing a bit-flip $m_j \rightarrow -m_j$ we change the sign of any pair of neighboring correlators. For instance, performing a bit-flip $m_1 \rightarrow -m_1$ on the box $(-1,1,1,1,...,1,1)$ produces another extremal contextual box $(1,-1,1,1,...,1,1)$. Consequently, performing a bit-flip on each consecutive $m_j$, we can generate all extremal contextual boxes with exactly one correlator equal to $-1$. Now, given a box $(1,1,1,1,...,1,-1)$ we again perform a bit-flip $m_1 \rightarrow -m_1$ which produces a box $(-1,-1,1,1,...,1,-1)$ from which we can generate all the boxes with 3 correlators equal to $-1$. In so doing we can generate all $2^{n-1}$ extremal contextual boxes (all those with odd number of correlators equal to $-1$) by a contextuality preserving operation. We see then that $\B^n$ satisfies vertex-equivalence property. Moreover, $X_{\max}$ does not change under bit-flip of outputs of some of the observables, hence assumptions of
Proposition \ref{ce4} are satisfied. In consequence the bound (\ref{eq:xmax-chain-bound}) is true.
\blacksquare

\section{A bound on distillable resource in box theories}

In this section we consider a scenario analogous to the scenario of distillation of entanglement in the entanglement theory. Namely, we assume that $n$ copies of a box $B$ are provided to a party (or parties in case of non-locality). We distinguish a target box $B_v^{T}$ which is valuable,
approximation of which one wants to ``distill'' out of $n$ copies of $B$. We demand that the distilling operations satisfy the axiom (O2), i.e., do not create a valuable box out of non-valuable ones. We define distillable resource $D(B_v^T|B)$ for a box $B$, in analogous manner to definition of distillable entanglement \cite{dist1,dist2} as the highest ratio
of number $k$ which is such that the output of distillation protocol approximates $[B_v^{T}]^{\otimes k}$, to the number of used boxes $B$ which is $n$, in asymptotic limit of large $n$. The main result of this section states, that the so called {\it regularized} measure of resource $X$, for which $X$ satisfies the axioms (M2) of monotonicity and (M4) asymptotic continuity, up to a constant factor $X(B_v^{T})$ is an upper bound to distillable resource $D(B_v^T|B)$:
\be
X^{\infty}(B) \geq X(B_v^T)D(B_v^T|B),
\label{eq:dist-bound}
\ee
where $X^{\infty} = \lim_{n \rightarrow \infty} {X(B^{\otimes n}) \over n}$ is the regularized measure $X$.

\subsection{Proof of the upper bound}

In this section we prove inequality (\ref{eq:dist-bound}) under some assumptions on measure $X$ and a target box $B_v^T$. We begin with definition of {\it rate of distillability of a box $B_v^{T}$ from a box $B$}, denoted as $D(B_v^{T}|B)$.

{\definition For a box $B \in \B$, consider a sequence $\Lambda_n $ of operations satisfying the axioms (O1) and (O2), such that $\Lambda_n(B^{\otimes n}) = B_n$. The set ${\cal D} =\{ \Lambda_n\}$ is called
a protocol distilling a target box $B_v^T$ from $B$, if
\be
\lim_{n\rightarrow \infty} ||B_n - [B_v^T]^{\otimes k_n}|| = 0.
\ee
For a given distillation protocol $\cal D$, its rate is given by
\be
r({\cal D}) = \limsup_{n\rightarrow \infty} {k_n \over n}.
\ee
The rate of distillability of the box $B_v^{T}$ from a box $B$ is given by:
\be
D(B_v^T|B) = \sup_{\cal D} r(\cal{D}).
\ee
}

We can proceed to show the main result of this section, which state that asymptotically continuous and monotonous measure of resource is, up to a constant factor, an upper bound on $D(B_v^T|B)$, as it is stated in proposition below.

{\proposition  Let $B_v^T \in \B_v$ be some target box. Let $X$ be a measure which satisfies the axioms of (M2) monotonicity, (M4) asymptotic continuity, and also let $X$ be superadditive on $B_v^T$. Then we have:
\be
D(B_v^{T}|B)X(B_v^{T}) \leq X^{\infty}(B).
\label{eq:prop-main-ineq}
\ee
\label{prop:dist-bound}
}

{\it Proof}.
Consider $n$ copies of a box $B$. The purpose is to distill the largest number of (approximate) copies of target boxes $B_v^{T}$. Let us fix $\delta >0$. Then there exists a protocol
${\cal D} = \{\Lambda_n\}$ such that $r({\cal D})> D(B_v^{T}|B) - \delta$. It follows also that for sufficiently large $n$
\be
\Lambda_n(B^{\ot n}) = B_n,
\label{eq:BtoPR}
\ee
such that there holds $||B_n - {B_v^{T}}^{\ot k_n}||_D\leq \epsilon_n$, where $0<\epsilon_n \rightarrow 0$ with $n \rightarrow \infty$.
Then we have the following chain of inequalities:
\ben
X(B^{\ot n}) &\geq& X(\Lambda_n(B^{\ot n})) \nonumber \\
&=& X(B_n) \nonumber \\
&\geq& X({[B_v^{T}]}^{\ot k_n}) - f(\epsilon_n)  \nonumber \\
&\geq& k_n X(B_v^{T}) - f(\epsilon_n),
\een
where the first inequality holds by the axiom of (M2) monotonicity of $X$ under operations $\Lambda_n$ satisfying the axiom (O2). The first equality is by (\ref{eq:BtoPR}), the second inequality is by asymptotic continuity of $X$, where $f(.)$ is some continuous function. The last equality is by the assumption of superadditivity of $X$ on a box $[B_v^T]^{\otimes k_n}$ (see in this context Theorem 9 of \cite{context-measures}).

If we now divide the first and the last term of the above chain of (in)equalities by $n$, we have:
\be
{X(B^{\ot n})\over n} \geq X(B_v^{T}){k_n\over n} - {f(\epsilon_n)\over n}.
\ee

In fact, the left-hand side of the above inequality approaches $X^{\infty}$ in the limit $n \rightarrow \infty$ and, by continuity of $f$, the right-hand side approaches $X(B_v^{T})[D(B_v^{T}|B) -\delta]$, as it was expected.
Since $\delta$ was arbitrary, taking the limit of $\delta \rightarrow 0$ proves inequality (\ref{eq:prop-main-ineq}).
\blacksquare

We have now the following remark:

{\rem If additionally a measure $X$ is subadditive, there is $X \geq X^{\infty}$, hence by the above proposition, we obtain:
\be\label{ogran}
X(B) \geq X(B_v^T)D(B_v^T|B).
\ee
}

\subsection{Protocol of distillation of contextuality}

Below we consider a particular example of a distillation protocol of contextual resources. We will consider a distillation protocol which uses two copies of weakly contextual resource and transform them into a single copy of more contextual resource. We will show this by a specific indicator of contextuality $\beta$ (which quantifies the violation of a specific contextual inequality) rather than a measure of contextuality $X_{\textrm{max}}$ which is uneasy to calculate for boxes used in the distillation process. The distillation protocol itself is easily implementable in the experiment, as it involves only postprocessing of the data (measurement outcomes).

First let us describe in detail contextual resources from which we distill more valuable resource.

An {\bf XOR-box} is a consistent box $\{p(a_1,a_2,..,a_m|c_i)\}$ such that for each input $c_i$ and binary outputs $a_1,...,a_m\in \{0,1 \}$, either ($p^{od}$: odd context)
\ben
 p^{od}(a_1,..,a_m|c_i)=
  \begin{cases}
   1\over{2^{m-1}} & \forall \oplus_j^m a_j=1 \\
   0       & \text{otherwise,}\\
  \end{cases}
\een
or ($p^e$: even context)
\ben
 p^e(a_1,..,a_m|c_i)=
  \begin{cases}
   1\over{2^{m-1}} & \forall \oplus_j^m a_j=0 \\
   0       & \text{otherwise.}\\
  \end{cases}
\een
Note that an XOR-box can in principle be contextual or noncontextual. For the rest of this section we assume that as an XOR-box we mean only contextual box, of which the examples are presented in Ref. \cite{context-measures} under the name of PM-box, M-box, CH-box.

A {\bf correlated box} is a box such that for each input $c_i$ and binary outputs $a_1,...,a_m\in \{0,1 \}$ there is
\ben
 p(a_1,..,a_m|c_i)=
  \begin{cases}
   1\over{2^{m-1}} & \forall \oplus_j^m a_j=0 \\
   0       & \text{otherwise.}\\
  \end{cases}
\een
All correlated boxes which correspond to a hypergraph of a certain XOR-box are noncontextual: a correlated box can be obtained from a single joint probability distribution $p(\mathcal{M})$ which is decomposable into probabilistic points with the property $\oplus_j^m a_j=0$ for all $c_i$.

The protocol of distillation that we will employ originally was used in \cite{dist-nonl} for distilling nonlocal resources, whereas in terms of contextual resources it works as follows: on inputs $c_i$ for two copies of a resources one inputs the same number and then receives outputs $(a_1,...,a_m)$ and $(a'_1,...,a'_m)$, respectively. Then one compute the final output as $(a_1\oplus a'_1,...,a_m\oplus a'_m)$. This procedure we will call a node wise XOR operation.

Furthermore, we will use the parameter $\beta$ (for a definition see Ref. \cite{context-measures}) as a contextuality indicator of a given box, by means of violating the contextual inequality: if we denote $B^*$ as a reference extremal isotropic XOR-box, then for any contextual box $B$ the contextual inequality
\ben\label{conineq}
\beta_{B^*}(B) \leq n-1
\een
($n$ - the number of contexts) is violated.
In the following theorem we will show that by performing XOR operation on two copies of a contextual box one can concentrate the contextuality content in terms of increasing the value $\beta$.

{\theorem
Let $\Lambda:\mathcal{B}_v \otimes \mathcal{B}_v \rightarrow \mathcal{B}_v$
be a linear and non-contextuality preserving node wise operating map.
There exist a map $\Lambda$ such that
\be
\beta_{B_x}(\Lambda(B_v^{\otimes2})) > \beta_{B_x}(B_v),
\ee
where $B_v \in \mathcal{B}_v$, and $\beta_{B_x}(B_v)=2^{m-1}\<B_x|B_v\>$, where $B_x$ is an extremal isotropic XOR-box.
}

The theorem says that there exist a non-contextuality preserving map which can be used for distillation of contextuality. Indeed, this result holds for any even number of copies of a box $B_v$. Below we will prove the theorem for $\Lambda$ being an XOR operation. Then we will show, that a node wise XOR operation is noncontextuality preserving, i.e., it cannot distill a contextual resource from noncontextual boxes $B\in \mathcal{B}_{nv}$.

{\it Proof.}
Let $\Lambda_{\textrm{XOR}}$ be the node wise XOR operation, i.e., for any context $c_i$ it acts as the following
\be
p(a_1,..a_m|c_i)\otimes p(a_1',..,a_m'|c_i)\overset{\Lambda_{\textrm{XOR}}}{\rightarrow} p(a_1\oplus a_1',..a_m\oplus a_m'|c_i).
\ee
Consider probability distributions, $p^{od},p^e$, which constitute a XOR-box. Let us check how $\Lambda_{\textrm{XOR}}$ acts on different compositions of $p^{od}$ and $p^e$:
\ben
p^e\otimes p^e &=& p(\oplus_j^m a_j=0|c_i)\otimes p(\oplus_j^m a_j'=0|c_i) \nonumber \\
 &\overset{\Lambda_{\textrm{XOR}}}{\rightarrow}& p(\oplus_j^m (a_j\oplus a_j')=0|c_i) \nonumber \\
 &=& p^e,
\een
where we used a simple identity $(\oplus_j^m a_j) \oplus(\oplus_j^m a_j') = \oplus_j^m (a_j\oplus a_j')$.
Similarly one can show the following
\ben \label{eq:eve-odd}
p^e\otimes p^{od} &\overset{\Lambda_{\textrm{XOR}}}{\rightarrow}& p^{od} \nonumber \\
p^{od}\otimes p^e &\overset{\Lambda_{\textrm{XOR}}}{\rightarrow}& p^{od} \\
p^{od}\otimes p^{od} &\overset{\Lambda_{\textrm{XOR}}}{\rightarrow}& p^e. \nonumber
\een

Consider now a box $B$ defined as a linear combination of the extremal isotropic XOR-box $B_x$ and a correlated box $B_c$:
\ben
B= \alpha B_x + (1-\alpha)B_c.
\een
It is easy to verify that
\ben\label{eq:betaB}
\beta_{B_x}(B)&=&2^{m-1}\<B_x|B\>  \nonumber \\
&=& 2^{m-1}\Big(\alpha \<B_x|B_x\>+(1-\alpha)\<B_x|B_c\>\Big) \nonumber \\
&=& (n-1+\alpha),
\een
since $\<B_x|B_x\>=n/2^{m-1}$ and $\<B_x|B_c\>=(n-1)/2^{m-1}$.
We see that the contextual inequality \eqref{conineq} is violated for any $\alpha \in (0,1]$, therefore $B\in \mathcal{B}_v$ except $\alpha=0$.

For two copies of the box $B$ we have
\ben \label{eq:2copy}
B^{\otimes 2}&=& \alpha^2 B_x^{\otimes 2}+(1-\alpha)^2 B_c^{\otimes 2} \nonumber \\ &+&\alpha(1-\alpha)(B_x\otimes B_c+B_c\otimes B_x),
\een
and after node wise XOR operation
\ben \label{eq:beta_chk}
\lefteqn{\beta_{B_x}(\Lambda_{\textrm{XOR}}(B^{\otimes2}))=2^{m-1}\<B_x|\Lambda_{\textrm{XOR}}(B^{\otimes2})\>} \nonumber \\
&&=2^{m-1}\Big(\alpha^2 \<B_x|\Lambda_{\textrm{XOR}}(B_x^{\otimes 2})\>+ (1-\alpha)^2\<B_x|\Lambda_{\textrm{XOR}}(B_c^{\otimes 2})\> \nonumber \\
&&+\alpha(1-\alpha)\<B_x|\Lambda_{\textrm{XOR}}(B_x\otimes B_c+B_c\otimes B_x)\> \Big).
\een
Taking into account \eqref{eq:eve-odd} one can show that
\ben
\<B_x|\Lambda_{\textrm{XOR}}(B_x^{\otimes 2})\>&=& (n-1) \<p^e|p^e\>+\<p^{od}|p^e\> \nonumber \\
 &=&(n-1)/2^{m-1}, \\
\<B_x|\Lambda_{\textrm{XOR}}(B_c^{\otimes 2})\>&=&(n-1) \<p^e|p^e\>+\<p^{od}|p^e\> \nonumber \\
 &=&(n-1)/2^{m-1},
\een
and
\ben
\lefteqn{\<B_x|\Lambda_{\textrm{XOR}}(B_x\otimes B_c+B_c\otimes B_x)\>} \nonumber \\
&&=2(n-1) \<p^e|p^e\>+2\<p^{od}|p^{od}\> =2n/2^{m-1},
\een
since for single-context probability vectors $\<p^e|p^e\>=1/2^{m-1}$ and $\<p^{od}|p^e\>=0$.

Inserting these values into equation (\ref{eq:beta_chk}) we get,
\be \label{eq:betaB2}
\beta_{B_x}(\Lambda_{\textrm{XOR}}(B^{\otimes2}))=(\alpha^2+(1-\alpha)^2)(n-1)+2\alpha(1-\alpha)n.
\ee
Then by comparing (\ref{eq:betaB}) with (\ref{eq:betaB2}) one can see that for $0<\alpha < 1/2$ we obtain
\be
\beta_{B_x}(\Lambda_{\textrm{XOR}}(B^{\otimes2})) > \beta_{B_x}(B).
\ee
$\blacksquare$

Note: we conjuncture that node wise XOR operation is the only operation which results in distillation.

We will now show that node wise XOR operation is indeed noncontextuality preserving, i.e., it satisfies the axiom (O2). Suppose that we aim to distill contextuality from noncontextual boxes $B$ and $B'$. We need to show that the box $B''=\Lambda_{\textrm{XOR}}(B\otimes B')$ is also noncontextual.
Now, since the box $B= \{p(a_1,...,a_m|c_i) \}$ ($B'= \{p'(a'_1,...,a'_m|c_i) \}$) is noncontextual, then there exist a joint probability distribution $p(a_1,a_2,...)$ ($p'(a'_1,a'_2,...)$) for all observables in $\mathcal{M}$. Denote $\lambda$ ($\lambda'$) as a string of $2^{|\mathcal{M}|}$ outputs $(a_1,a_2,...)$ ($(a'_1,a'_2,...)$), so that $p(a_1,a_2,...)$ ($p'(a'_1,a'_2,...)$) is a linear combination of deterministic points indexed by $\lambda$ ($\lambda'$), each with probability $p(\lambda)$ ($p'(\lambda')$). Note that a node wise XOR operation is a map $\Lambda_{\textrm{XOR}}: \{\lambda\}\times \{\lambda' \} \rightarrow \{ \lambda'' \}$, where the string of outputs $\lambda''$ are defined by
\ben\label{compos}
(a''_1,a''_2,...)= (a_1 \oplus a'_1,a_2 \oplus a'_2,...).
\een
The box $B''$ is then a linear combination of deterministic points indexed by $\lambda''$, each with probability
\be
p''(a''_1,a''_2,...)=\sum_{\{\lambda\} \times \{\lambda'\}|{\oplus}} p(a_1,a_2,...)p'(a'_1,a'_2,...),
\ee
where the above sum is over all composition of strings $\lambda$ and $\lambda'$ such that \eqref{compos} holds. For example in case of $|\mathcal{M}|=2$ we would have e.g. for a string $(a''_1=0,a''_2=1)$
\ben
\lefteqn{p''(01)} \\
&&=p(00)p'(01)+p(01)p'(00)+p(10)p'(11)+p(11)p'(10). \nonumber
\een
Note also that $p''(a''_1,a''_2,...)$ forms a well defined probability distribution, because summing all probabilities
\ben
\sum_{\{\lambda''\}} p''(a''_1,a''_2,...)&=&\sum_{\{\lambda\} \times \{\lambda'\}} p(a_1,a_2,...)p'(a'_1,a'_2,...) \nonumber \\
&=&\sum_{\{\lambda\}} \sum_{\{\lambda'\}} p(a_1,a_2,...)p'(a'_1,a'_2,...) \nonumber \\
&=& 1.
\een
The first equality is based on the observation that the inverse image of the map $\Lambda_{\textrm{XOR}}$ for all elements in $\{\lambda''\}$ results in disjoint partitions of the entire product set $\{\lambda\} \times \{\lambda'\}$.
Thus we have shown that the box $B''$ is noncontextual since there exist a joint probability distribution $p''(a''_1,a''_2,...)$ which defines $B''$.
Similarly one can show that any node wise operation is also noncontextuality preserving.

\subsection{Towards application of Proposition \ref{prop:dist-bound}}

We can pass now to consider for which resources and measures the assumptions of the above proposition are satisfied. We consider bipartite scenario
with non-local correlations as a resource, as that for contextuality follows the same lines, and faces the same problems.

\begin{enumerate}
\item (Possible bound via $X_{\max}$) Consider $B_v^{T}$ to be $PR$-box, and a measure $X$ to be $X_{\max}$. Then $X_{\max}$ is additive on $B_v^{T}$ (see \cite{context-measures}).
By Theorem \ref{thm:main-as-cont}, $X_{\max}$ is also asymptotically continuous. We can consider distillation protocol via restricted set of operations, namely the {\it wirings}  \cite{Allcock-wires}, as for suitably defined non-valuable, i.e., local boxes  \cite{Joshi-non-broad} (see easier formulation \cite{moja-distinguishing}) transforms local boxes into local ones, hence satisfy the axiom (O2) (see Appendix \ref{Appwir}). However, one would need a proof that $X_{\max}$ does not increase under wirings, which we leave as an open question. It is easy to check that for isotropic boxes $PR_{\alpha} = \alpha B_{000} + (1-\alpha)B_{001}$  (for a formal definition see \cite{context-measures}) this bound would be nontrivial in the whole range of $\alpha \in (3/4,1]$.

\item (On bound via $X_{\uu}$) Similarly, as for $X_{\max}$, the measure $X_{\uu}$ is additive on $PR$-boxes, and is asymptotically continuous via proof analogous
to that of Theorem \ref{thm:main-as-cont}. However, it is definitely not
monotonous under general operations which satisfy the axiom (O2), i.e., those that transform local boxes into local ones. This is because it can increase under partial trace.
Indeed, consider a box with a hypergraph $G$ equal to a direct sum of two hypergraphs $G_1\oplus G_2$, such that $G_2$ has two vertices connected by a single edge,
and the context corresponding to this edge is locally with Alice. Let also $G_1$ be a hypergraph of a non-local box. Let now the parties have a box $B$ corresponding
to $G$, which is $PR$-box, and a local box with Alice called $L$ so that the box $B$ equals $PR\oplus L$. By Theorem 8 of \cite{context-measures} there is
\be
X_{\uu}(PR\oplus L) = {4\over 5} X_{\uu}(PR) + {1\over 5}X_{\uu}(L).
\ee
Now, since $X_{\uu}(L)=0$ as this box is local, we have that $X_{\uu}(PR\oplus L) < X_{\uu}(PR)$, hence, by removing/adding $L$ one can increase/decrease the value of $X_{\uu}$.
Let us note here, that $X_{\max}$ does not suffer from the same problem, as
\ben
X_{\max}(PR\oplus L) &=& \max \{X_{\max}(PR),X_{\max}(L)\} \nonumber \\
&=& X_{\max}(PR),
\een
so that adding or removing a local box does not change value of $X_{\max}$.
Despite the fact that $X_{\uu}$ is not monotonous under locality preserving operations, monotonicity under wirings is still possible for it, which we also leave as an open problem.

\end{enumerate}

It is worth mentioning that while considering the measure of contextuality $X_{\uu}$ we observe that it is a normalized version of nonlocality quantifier as referred in Ref. \cite{Vin14}. Notice, however, that although the unnormalized statistical distance measure of nonlocality, given by infimum over local distributions may increase under local transformations (in particular enlarging the number of inputs of a box \cite{Vin14}), it is not necessarily so when the number of added new inputs are properly accounted. Thus, a normalized measure of contextuality ($X_{\uu}$ as well as $X_{\max}$) prevents the increase of relative entropy while trivial expansion of the number of contexts takes place.


\section{Conclusions}

Using an axiomatic approach common to resource theories, we
have developed the theories of contextuality, and its most
celebrated example, which is non-locality. Crucially from the experimental point
of view, we have studied the axiom of asymptotic continuity,
and proved that recently established measure of
contextuality --- the {\it relative entropy of contextuality} \cite{context-measures} obeys that axiom. We thereby have
showed that for an experimental setup which
produces an imperfect box $B'$, close to the intended box
$B$, the amount of contextuality measured by the relative entropy of contextuality $X(B')$ cannot differ from $X(B)$ by more than
the distance $||B - B'|| $ with a factor depending only logarithmically on the dimension
of the boxes.

We have also considered a general measure of resource $X$,
with properties satisfying three proposed axioms/properties: of faithfulness, local invariance
and convexity. We have focused on boxes $B$ from the polytope satisfying vertex equivalence property, i.e.,
which is such that all its contextual vertices are reversibly exchangeable into each other.
We have shown that in such polytopes the measure $X$ is  upper bounded by the measure called the {\it cost of
the resource} $C(B)$ with a multiplication factor  $X(E_v)$ for some extremal valuable box $E_v$.
Interestingly, due to this factor, we were able to bound an {\it extensive measure} (which grows linearly with number of copies), by a {\it non-extensive one} (which takes values in $[0,1]$ on any box irrespective of its dimension).
The mentioned bound is linear function of the box. It would be interesting to
find a non-linear one, which is more tight and still easily computable.
We have supported the latter results by two examples of its application:
for bipartite boxes with binary inputs and outputs, as
well as for the boxes related to contextual chain box.
Analogous, but weaker upper bound holds in the case of the polytopes $\B$, which do not satisfy
the vertex equivalence property of $\B$.

We have studied a distillation protocol of a valuable
target box $B^T_v$ from many copies of some input boxes $B$, and in full analogy with theory of
entanglement measures, we have provided an upper bound on the rate of distillability of the resource
$D(B^T_v |B)$. It is expressed by a measure of resource $X$ which
satisfies another two proposed axioms: of monotonicity under allowed class of operations,
asymptotic continuity and superadditivity on target boxes: $X([B^T_v]^{\otimes k}) \geq k X(B^T_v)$.
From our investigation we can conclude, that the relative entropy of contextuality for bipartite boxes with
two binary inputs and outputs may be an upper bound on distillable non-locality in the form of the Popescu-Rohrlich boxes.
The only fact which needs to hold for the latter to be true, is the non-increasing of this measure
under wirings. We leave this remaining question as an open problem.

Finally, checking whether other measures of contextuality or non-locality such as, e.g., \cite{Svozil-context,Kleinmann-memory-cost} satisfy proposed axioms, would be vital for their further use, and it would be also interesting to find new ones which satisfy the axioms by definition.

{\bf Note added}: While finishing this manuscript, we became aware of the results of \cite{Coke-resource}. It seems that our results can be set in a more general framework of general resource theories formulated there in more abstract language.

\begin{acknowledgments}
This work is supported by ERC Advanced Grant QOLAPS, and the Polish Ministry of Science and Higher Education
Grant no. IdP2011 000361.  PJ is supported by grant MPD/2009-3/4 from Foundation for Polish Science. Part of this work was done in National Quantum Information Centre of Gda{\'n}sk.
\end{acknowledgments}

\begin{appendix}

\section{Formal proof of Lemma \ref{lem:inf-ascont}}\label{APP1}

Here we give the formal proof of Lemma \ref{lem:inf-ascont}.

{\it Proof.}
By definition of infimum, for any $\delta_n > 0$ there exists a sequence $\rho_n$ such that
\be
f^*_{T}:=\inf_{\rho \in T} f(\rho) \leq f(\rho_n) \leq f^*_{T} + \delta_n.
\label{eq:def_inf}
\ee
Moreover, by the assumption (\ref{eq:assumption1}) there exists a sequence $\sigma_n$, such that
\be
f(\rho_n) - g(\delta) \leq f(\sigma_n) \leq f(\rho_n) + g(\delta).
\ee
Combining the above two sequences of inequalities we obtain:
\be
f^*_{T} - g(\delta) \leq f(\sigma_n) \leq f^*_{T} + \delta_n + g(\delta).
\ee

Now, there exists $n_0$ such that for every $n\geq n_0$ there holds $\delta > \delta_n$, and hence
\be
f^*_{T} - g(\delta) -\delta \leq f(\sigma_n) \leq f^*_{T} + \delta + g(\delta).
\label{eq:inequality_inf}
\ee
This means that we obtained a sequence $f(\sigma_n)$ which is bounded (we use here the fact that the infima are bounded), and by Bolzano-Weierstrass theorem
there exists a subsequence $n_k$, such that $f(\sigma_{n_k})$ has a limit. Thus we have in particular:
\be
\lim_{n \rightarrow \infty} f(\sigma_{n_k}) - f^*_{T} \leq g(\delta) + \delta.
\ee
Since by definition $f^*_{T'} := \inf_{T'} f(\sigma) = \lim_{n \rightarrow \infty} f(\sigma_{n_i})$ for some sequence $\{\sigma_{n_i}\}$, we have from the above inequality, that $\{\sigma_{n_k} \}$ may be suboptimal (the infimum over a set is the infimum of the set of limits of sequences from this set), hence
\be
f^*_{T'} - f^*_{T} \leq g(\delta) + \delta.
\ee
Analogously, exchanging $T$ and $T'$ we can arrive at
\be
f^*_{T} - f^*_{T'} \leq g(\delta) + \delta,
\ee
which proves the thesis for infima. The proof for supremum goes analogously, with only change of inequalities to opposite and signs in front of $\delta_n$ in (\ref{eq:def_inf}), which leads us exactly to the expression (\ref{eq:inequality_inf}), but for the supremum. The rest of the proof goes symmetrically, hence we skip it.\blacksquare

\section{Convexity of $I_{\max}$}\label{app:conv}

In this section we will present an explicit proof of another property of the measure of contextuality, which is its convexity. This property was used in Ref. \cite{PR} (see Eq. (2)), but without formal proof.
We will first prove convexity of $I_{p(c)}$ and then using the definition of supremum we will show convexity of the measure $I_{\max}$.
Note, that the convexity of the mutual information of contextuality $I_{\max}$ means that the relative entropy of contextuality $X_{\max}$ is also convex because of equivalence of the two measures.

Let us denote $B_{\mix}$ as a convex combination of boxes:
\ben
B_{\mix} = \sum_i p_i B_i,
\een
where $B_i$ are not necessarily extremal (or deterministic) boxes. Then, by definition of a box we have:
\ben
p_{B_{\mix}}(\lambda_c) = \sum_i p_i p_{B_{i}}(\lambda_c).
\een
We now have the following:
\ben\label{jeden}
\lefteqn{I_{p(c)}(\sum_i p_i B_i)} \nonumber \\
&=& I_{p(c)}(B_{\mix}) \nonumber \\
&=& \min_{p(\lambda)}\sum_c p(c)D(p_{B_{\mix}}(\lambda_c)||p(\lambda_c)) \nonumber \\
&\leq& \sum_c p(c)D(p_{B_{\mix}}(\lambda_c)||\sum_i p_i p^{i*}(\lambda_c)),
\een
where $p^{i*}(\lambda_c)$ is a marginal distribution obtained from a joint probability distribution $p^{i*}(\lambda)$ optimal for a particular box $B_i$. The above inequality comes from the fact that the distribution $p^*(\lambda) = \sum_i p_i p^{i*}(\lambda)$ not necessarily gives a desired minimum over all distributions $p(\lambda)$.
Furthermore, we have
\ben\label{dwa}
\lefteqn{\sum_c p(c)D(\sum_i p_i p_{B_i}(\lambda_c)||\sum_i p_i p^{i*}(\lambda_c)) } \nonumber \\
&\leq& \sum_i p_i \sum_c p(c) D( p_{B_i}(\lambda_c)|| p^{i*}(\lambda_c)) \nonumber \\
&=& \sum_i p_i I_{p(c)}(B_i),
\een
where the inequality comes from joint convexity of relative entropy distance for each $c$, while for the last equality we utilized the optimality of $p^{i*}(\lambda)$ for each box $B_i$. Using the results \eqref{jeden} and \eqref{dwa} we arrive at
\be
I_{p(c)}(\sum_i p_i B_i) \leq\sum_i p_i I_{p(c)}(B_i).
\ee

Now, by the definition of supremum, for any $\delta_n >0$ there exists a distribution $p_n(c)$, such that
\be
I_{\max}(\sum_i p_i B_i) \leq I_{p_n(c)}(\sum_i p_i B_i) + \delta_n,
\ee
hence, by convexity of $I_{p_n(c)}$ we have
\be
I_{\max}(\sum_i p_i B_i) \leq \sum_i p_i I_{p_n(c)}(B_i) + \delta_n.
\ee
Notice that for each $i$ the definition of $I_{\max}$ assures that $I_{p_n(c)}(B_i) \leq I_{\max}(B_i)$. Thus
\be
I_{\max}(\sum_i p_i B_i) \leq \sum_i p_i I_{\max}(B_i) + \delta_n,
\ee
and because $\delta_n$ can be arbitrarily small, we obtain the desired convexity of $I_{\max}$.

\section{Locally performed wirings satisfies the axiom (O2)}\label{Appwir}

In this section we present a formal proof of the fact that if Alice and Bob have access to $n$ boxes, such that the collection of the latter admits a local hidden variable model, then by means of locally performed wirings \cite{Allcock-wires} one cannot transform the collection of boxes into a valuable (nonlocal) resource shared by the two parties.

Consider then a collection of $n$ boxes shared by Alice and Bob which admits a local hidden variable model with respect to both parties
\ben\label{belokal}
B_L^n = \sum_\lambda p_\lambda p^{(\lambda)}(\textbf{a}|\textbf{x})\otimes p^{(\lambda)}(\textbf{b}|\textbf{y}),
\een
for some probability distribution $\{ p_\lambda \}$, where $\textbf{a} = (a_1,...,a_m)$ ($\textbf{b} = (b_1,...,b_m)$) is the vector of Alice's (Bob's) outputs when one of the input from $\textbf{x} = (x_1,...,x_n)$ ($\textbf{y} = (y_1,...,y_n)$) is chosen.

We will assume, that {\it locally} the distribution of the collection of boxes as seen by one party (e.g. Alice), $p(\textbf{a}|\textbf{x})$, is nonsignalling \cite{moja-distinguishing}, i.e. the following conditions are satisfied
\be\label{nsduze}
\forall_{1\leq i\leq m,\textbf{a}^{\neq i},\textbf{x}^{\neq i},x_i,x_i'}\,\, \sum_{a_i} p(\textbf{a}|\textbf{x}^{\neq i},x_i) = \sum_{a_i} p(\textbf{a}|\textbf{x}^{\neq i},x_i'),
\ee
and analogously for Bob. Note that the nonsignalling conditions given above implies nonsignalling with respect to all subsets of inputs, i.e., marginal distribution of the outputs $\textbf{a}^{\neq i,j,...}$ does not depend on changing the inputs $\textbf{x}^{\neq i,j,...}$ \cite{hangi}.

Consider now the partition of constituent boxes $A_1:A_2:B$, where $A_1 \equiv \{x_1,...,x_k\},A_2\equiv\{x_{k+1},...,x_n\},B\equiv\{y_{1},...,y_n\}$ for an arbitrary $1\leq k \leq n-1$. As it was shown in Ref. \cite{toble}, the locality in the partition $A_1,A_2:B$ may not be preserved when the subsystems $A_1$ and $A_2$ cooperate, i.e., when they perform a suitable wiring. This happens when no constraints are imposed on the distribution $p^{(\lambda)}(\textbf{a}|\textbf{x})$ in the decomposition \eqref{belokal}. However, when the local distribution admits nonsignaling conditions \label{nsduze}, then the operation of wiring of the subsystems $A_1$ and $A_2$ will not lead to emergence of signaling for the one-partite distribution $p(\textbf{a}|\textbf{x})$. Since such nonsignalling bilocal distributions (NSBL) constitute a closed set under wirings \cite{toble}, then we see that locally performed the operation of wirings will not produce a valuable (nonlocal) resource from useless (local) objects (see Supplemental Material of Ref. \cite{toble}, where the nonsignalling conditions (C2) need to be assumed).

\end{appendix}


\begin{thebibliography}{99}



\bibitem{KS67}
\bia{S. Kochen, E.P. Specker}{}{J. Math. Mech.}{17}{59}{1967}

\bibitem{Bell66}
\bia{J.S. Bell}{}{Rev. Mod. Phys.}{38}{447}{1966}

\bibitem{con1}
\bia{A. Cabello}{}{Phys. Rev. Lett.}{101}{210401}{2008}

\bibitem{con2}
\bia{A.A. Klyachko, M.A. Can, S. Binicio\u{g}lu, A.S. Shumovsky}{}{Phys. Rev. Lett.}{101}{20403}{2008}

\bibitem{con3}
\bia{R. Lapkiewicz, P. Li, C. Schaeff, N.K. Langford, S. Ramelow, M. Wie\'{s}niak, A. Zeilinger}{}{Nature}{474}{490}{2011}

\bibitem{con4}
\bia{Kirchmair {\it et al.}}{}{Nature}{460}{494}{2009}

\bibitem{raus}
R. Raussendorf, Phys. Rev. A \textbf{88}, 022322 (2013).

\bibitem{CAEGCXL14}
\bia{G. Ca\~{n}as, M. Arias, S. Etcheverry, E.S. G\'{o}mez, A. Cabello, G.B. Xavier, G. Lima}{}{Phys. Rev. Lett.}{113}{090404}{2014}

\bibitem{BellNL}
\bia{N. Brunner, D. Cavalcanti, S. Pironio, V. Scarani, S. Wehner}{}{Rev. Mod. Phys.}{86}{419}{2014}

\bibitem{Hor4}
\bia{R. Horodecki, P. Horodecki, M. Horodecki, K. Horodecki}{}{Rev. Mod. Phys.}{81}{865}{2009}

\bibitem{Vin14}
\bia{J.I. de Vicente}{}{J. Phys. A: Math. Theor.}{47}{424017}{2014}

\bibitem{ressteer}
R. Gallego, L. Aolita, arXiv:quant-ph/1409.5804.

\bibitem{fryc}
B. Coecke, T. Fritz, R.W. Spekkens, arXiv:quant-ph/1409.5531.

\bibitem{nat}
\bia{M. Howard, J.J. Wallman, V. Veitch, J. Emerson}{}{Nature}{510}{351}{2014}

\bibitem{CT}
\bia{M. Christandl, B. Toner}{}{J. Math. Phys.}{50}{042104}{2009}

\bibitem{context-measures}
A. Grudka, K. Horodecki, M. Horodecki, P. Horodecki, R. Horodecki, P. Joshi, W. K{\l}obus, A. W\'{o}jcik,
Phys. Rev. Lett. \textbf{112}, 120401 (2014).

\bibitem{Rastall85}
P. Rastall, Found. Phys. 15, 963 (1985).

\bibitem{PR94}
\bia{S. Popescu, D. Rohrlich}{}{Found. Phys.}{24}{379}{1994}

\bibitem{AQBCC13}
\bia{M. Ara\'{u}jo, M.T. Quintino, C. Budroni, M.T. Cunha, A. Cabello}{}{Phys. Rev. A}{88}{022118}{2013}



\bibitem{dist-nonl}
\bia{M. Forster, S. Winkler, S. Wolf}{}{Phys. Rev. Lett.}{102}{120401}{2009}



\bibitem{Fine}
\bia{A. Fine}{}{Phys. Rev. Lett.}{48}{291}{1982}

\bibitem{Barrett}
\bia{J. Barrett}{}{Phys. Rev. A}{75}{032304}{2007}

\bibitem{Allcock-wires}
J. Allcock, N. Brunner, N. Linden, S. Popescu, P. Skrzypczyk, T. Vertesi, Phys. Rev. A \textbf{80}, 062107 (2009), arXiv:0908.1496.

\bibitem{DGG}
W. van Dam, R.D. Gill, P.D. Gr\"{u}nwald, IEEE Trans. Inf. Theory \textbf{51}, 2812 (2005).

\bibitem{Synak-Horodecki}
B. Synak-Radtke, M. Horodecki,
J. Phys. A: Math. Gen. \textbf{39}, L423 (2006).

\bibitem{Alicki-Fannes}
R. Alicki and M. Fannes, J. Phys. A: Math. Gen \textbf{37}, L55
(2004), quant-ph/0312081.

\bibitem{PR}
\bia{R. Ramanathan, P. Horodecki}{}{Phys. Rev. Lett.}{112}{040404}{2014}

\bibitem{BC90}
\bia{S.L. Braunstein, C.M. Caves}{}{Ann. Phys.}{202}{22}{1990}

\bibitem{Weh}
\bia{S. Wehner}{}{Phys. Rev. A}{73}{022110}{2006}

\bibitem{dist1}
\bia{C.H. Bennett, H.J. Bernstein, S. Popescu, B. Schumacher}{}{Phys. Rev. A}{53}{2046}{1996}

\bibitem{dist2}
\bia{C.H. Bennett, D.P. DiVincenzo, J.A. Smolin, W.K. Wooters}{}{Phys. Rev. A}{54}{3824}{1996}

\bibitem{Joshi-non-broad}
\bia{P. Joshi, A. Grudka, K. Horodecki, M. Horodecki, P. Horodecki, R. Horodecki}{}{QIC}{13}{567}{2013}

\bibitem{moja-distinguishing}
K. Horodecki, arXiv:quant-ph/1401.4899.

\bibitem{toble}
\bia{R. Gallego, L.E. W\"{u}rflinger, A. Ac\'{i}n, M. Navascu\'{e}s}{}{Phys. Rev. Lett.}{109}{070401}{2012}

\bibitem{hangi}
E. H\"{a}nggi, R. Renner, S. Wolf, arXiv:quant-ph/0911.4171.

\bibitem{Coke-resource}
B. Coecke, T. Fritz, Robert W. Spekkens, arXiv:1409.5531

\bibitem{Svozil-context}
K. Svozil, arXiv:1103.3980


\bibitem{Kleinmann-memory-cost}
M. Kleinmann, O. G�hne, J. R. Portillo, J. Larsson, A. Cabello, New J. Phys. 13, 113011 (2011)


\end{thebibliography}

\end{document}